# Similarity-Detection and Localization

Terence Hwa[(1)] and Michael Lässig[(2)]
[(1)] *Physics Department, University of California at San Diego, La Jolla, CA 92093-0319*
[(2)] *Max-Planck-Institut für Kolloid- und Grenzflächenforschung, Kantstr. 55, 14513 Teltow, Germany*

The detection of similarities between long DNA and protein sequences is studied using concepts of statistical physics. It is shown that mutual similarities can be detected by sequence alignment methods only if their amount exceeds a threshold value. The onset of detection is a critical phase transition viewed as a localization-delocalization transition. The *fidelity* of the alignment is the order parameter of that transition; it leads to criteria to select optimal alignment parameters.

Evaluating the similarities between long strings of alphabet is a challenging task, which arises in many fields ranging from data processing to biology [1]. Standard applications involve the comparisons of copies of a message sequence blurred by an imperfect transmission or reproduction process. A particularly important example is evolution in biological systems, a process that mutates gene sequences in various ways including local *substitutions*, *insertions*, and *deletions*. Molecular biologists routinely compare newly sequenced genes to known ones in databases, a first step towards unveiling the structure and function of the new findings. This so-called *sequence alignment* is the most widely used mathematical/computational tool in molecular biology [2].

A (global) *alignment* of two sequences $\{P_i\}$ and $\{Q_j\}$ is defined as an ordered set of pairings $(P_i, Q_j)$ and of unpaired elements $(P_i, -)$ and $(-, Q_j)$ called gaps, each letter $P_i$ and $Q_j$ belonging to exactly one pairing or gap (see Fig. 1). The optimal alignment of the two sequences is determined by minimization of a cost or "energy" function $E$ favoring pairs of matching elements $(P_i = Q_j)$ over mismatches $(P_i \neq Q_j)$ and gaps. A simple and commonly used energy function is the sum over all matches, mismatches, and gaps of the alignment contributing the respective energies $-1$, $\mu > 0$, and $\delta > 0$ each [3]. Many more general energy functions have been discussed [2].

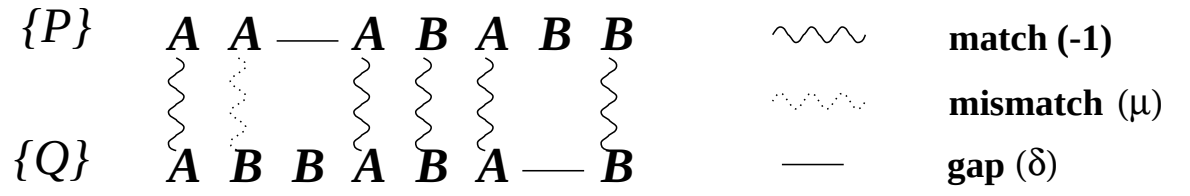

FIG. 1. One possible alignment of two binary sequences, $\{P_i\} = AAABABB$ and $\{Q_j\} = ABBABAB$, with six pairings (five matches, one mismatch) and two gaps.

Clearly, the optimal alignment depends strongly on these energy parameters, and so does its *fidelity*, i.e., the extent to which it captures mutual correlations between the sequences compared. In particular, if the evolution process involves random insertions and deletions, one has to allow for gaps in the optimal alignment so that mutually correlated regions of the sequences can actually align. For gene sequences, a more stringent criterion is the *biological relevance* of an alignment, that is, the extent to which the matched regions actually indicate functional similarities between different proteins. Finding alignment parameters that lead to high relevance is a difficult problem, which has not been solved systematically so far. For biological applications, "optimal" parameters for various types of genes are often deduced empirically from sequence pairs whose functional alignments are already known [4].

In this Letter, we apply ideas and methods of statistical physics to sequence alignment, introducing a different conceptual approach towards the parameter selection problem. Using simple stochastic models of evolution, we mutate an ancestor sequence to obtain daughter sequences $\{P_i\}$ and $\{Q_j\}$ with well-defined mutual correlations, namely, the ensemble of all pairs $(P_i = Q_j)$ of unmutated daughters of the same ancestor element. These sequences are then aligned using a given energy function $E$, and the fidelity of the optimal alignment is quantified as the fraction of correctly recovered such pairs $(P_i = Q_j)$. For long sequences, we find that the mutual correlations cannot be detected if their amount is below a threshold value. A critical transition separates this *low-similarity phase* of zero fidelity from the *high-similarity phase* where the fidelity is finite. (This phase transition is distinct from the "transition" to the so-called local alignment regime discussed in the literature [5].)

Our analysis of the phase transition is based on the known representation [6] of an alignment of two sequences $\{P_i\}$, $\{Q_j\}$ on the two-dimensional lattice of Fig. 2. The cells of this lattice are labeled by the index pairs $(i, j)$, or alternatively by the rotated coordinates $t = i + j$ and $r = i - j$. The bonds encode the adjacency of letters: the diagonal bond in cell $(i, j)$ represents the pairing $(P_i, Q_j)$; horizontal and vertical bonds correspond to gaps $(P_i, -)$ and $(-, Q_j)$, respectively. Thus any alignment maps onto a lattice path that is *directed* along the $t$ coordinate, i.e., given by a unique function $r(t)$. For daughter sequences generated from our stochastic evolution model, mutual similarities can be represented on the lattice by a path $R(t)$ joining all correlated pairs ($P_i = Q_j$). We call such a path the "target path". For long sequences, it induces a *morphological transition* on the optimal alignment path $r_0(t)$: in the low-similarity phase, this path is *super-diffusive* with typical fluctuations $\delta r(t) \equiv r_0(t) - R(t)$ scaling as $|\delta r(t)| \sim t^{2/3}$, while in the high-similarity phase, it is *localized* to the target path with finite fluctuations $|\delta r(t)| \sim \xi_\perp$; see Fig. 3. Hence, only in the high-similarity phase is the path $r_0(t)$ a faithful approximation of the target $R(t)$. The fidelity of the alignment is then simply given by the *overlap* of the two paths, i.e., by their number of intersections per unit of $t$. For long sequences, the overlap is proportional to the inverse localization length $\xi_\perp^{-1}$; maximization of this "order parameter" gives a numerically and analytically accessible criterion for the choice of alignment parameters.

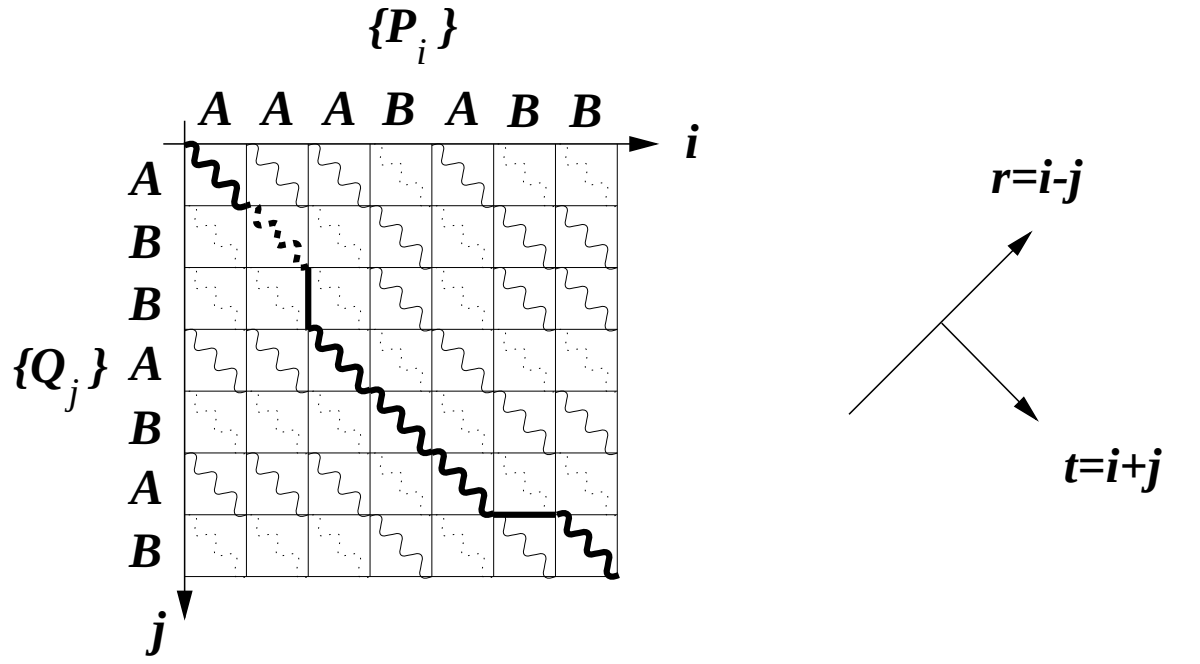

FIG. 2. The *directed* path (thick line) corresponding to the alignment in Fig. 1. Horizontal and vertical lattice bonds represent gaps, diagonal bonds represent pairings with bond energies $v_{r,t} = -(J \pm \Delta)$ for matches (full-wiggly lines) and mismatches (dashed-wiggly lines), respectively.

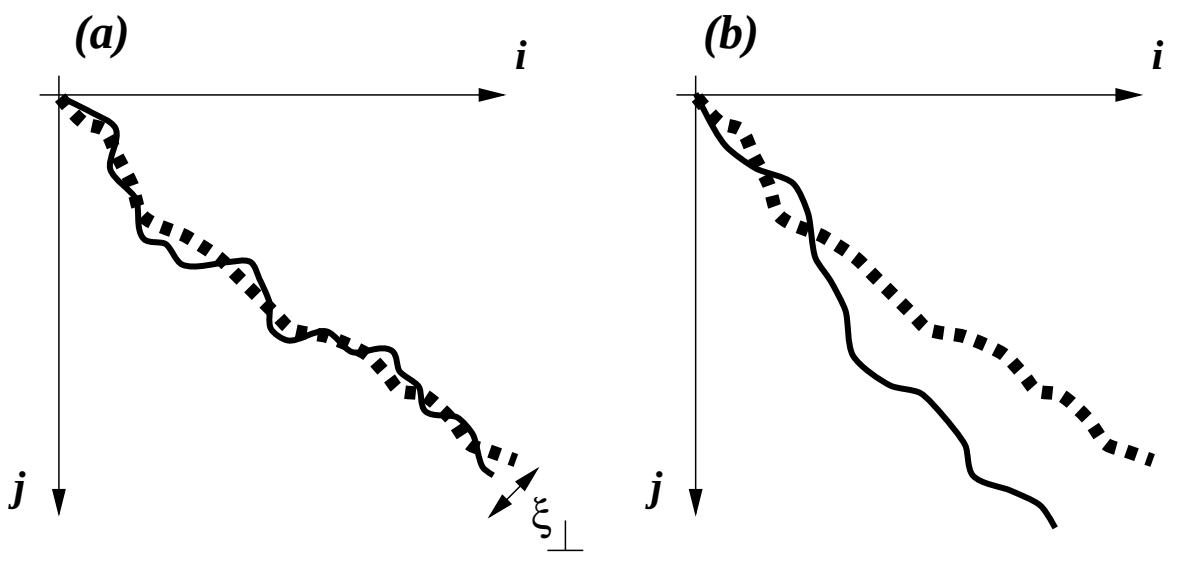

FIG. 3. Typical large-scale configurations of the optimal alignment path $r_0(t)$ for long sequences. (a) In the high-similarity phase, this path (full line) is *localized* to the target path $R(t)$ (dashed line) with typical transverse displacement of size $\xi_\perp$. (b) In the low-similarity phase, it fluctuates independently of the target path.

To establish these results, we restrict ourselves here to binary sequences (with elements $P_i, Q_j \in \{+1, -1\}$), and to the simple model introduced above with just two parameters $\mu$ and $\delta$ (generalizations are briefly discussed at the end of this Letter). With a convenient shift in the energy function, we write the total energy $E$ of any path $r(t)$ as the sum over all its diagonal bonds with bond energies $v_{r,t} \equiv -J - \Delta\, P_{i=(r+t)/2}\, Q_{j=(r-t)/2}$; horizontal and vertical bonds take the energy 0. The parameters $J \equiv 2\delta - (\mu - 1)/2$ and $\Delta = (\mu + 1)/2$ are the effective gap and mismatch costs; they characterize the stiffness of the alignment paths and the site-to-site variations of the pairing potential $v_{r,t}$, respectively. The lowest energy path $r_0(t)$ depends on $\mu$ and $\delta$ only via the ratio $g \equiv \Delta/J$. We shall restrict our analysis to the biologically interesting limit of $g < 1$ or $2\delta > \mu$ [8].

To model the behavior of typical alignment paths in the *low-similarity phase*, we take the sequences $\{P_i\}$ and $\{Q_j\}$ from an ensemble of unbiased *random sequences* (i.e., $\overline{P_i} = \overline{Q_j} = 0$, $\overline{P_i P_{i'}} = \delta_{i,i'}$, and $\overline{Q_j Q_{j'}} = \delta_{j,j'}$) with no mutual correlations, $\overline{P_i Q_j} = 0$. (Averages over this ensemble are denoted by overbars.) The pairing potential $v_{r,t}$ then becomes a *random* potential with average $\overline{v_{r,t}} = -J$ and variance

$$\begin{aligned} \overline{v_{r,t} v_{r',t'}} - J^2 &= \Delta^2\, \overline{P_{(r+t)/2} P_{(r'+t')/2} Q_{(r-t)/2} Q_{(r'-t')/2}} \\ &= \Delta^2 \delta_{r,r'} \delta_{t,t'}\,. \end{aligned} \tag{1}$$

It induces random fluctuations on the alignment paths.

The large-scale statistics of these fluctuations can be derived from the partition function of the alignment paths [7] in a path integral representation [9]. We show [8] that the continuum "action" of this path integral takes the form [10]

$$S = \int \mathrm{d}\tau \left[ \frac{\gamma}{2}\, \dot\rho^2 + \eta(\rho(\tau), \tau) + O(\dot\rho^4, \eta\dot\rho^2, \ldots) \right]. \tag{2}$$

It describes a "coarse-grained" alignment path $\rho(\tau)$ with a finite line tension $\gamma$ (and $\dot\rho \equiv \mathrm{d}\rho/\mathrm{d}\tau$) in a coarse-grained random potential $\eta(\rho, \tau)$ characterized by its second moment $\overline{\eta(\rho,\tau)\eta(\rho',\tau')} \sim \Delta^2 \delta(\rho - \rho')\delta(\tau - \tau')$. All other short-ranged moments [11], as well as the terms omitted in (2), are irrelevant variables. We conclude that the large-distance behavior of alignments is governed by the well-known universality class of a directed path in a two-dimensional *Gaussian* random potential [9]. Many properties of this universality class are known exactly. Typical fluctuations of the optimal path $r_0(t)$ are (asymptotically) super-diffusive, $\overline{(r_0(t) - r_0(t'))^2} \simeq A\, |t - t'|^{4/3}$, reflecting the tendency of the optimal alignment to use gaps to gain an excess number of matches over mismatches. This goes along with the variance $\overline{E_0^2(N)} - \overline{E_0(N)}^2 \simeq B\, N^{2/3}$ of the optimal energy $E_0(N)$ for paths of length $N \gg 1$. We have verified these properties numerically; an example is shown in Fig. 4. The nonuniversal amplitudes $A$ and $B$ have also been obtained. Scaling arguments, which are supported by our numerics, yield $A \simeq g^{4/3}$ and $B \simeq g^{4/3} J^2$ in the biologically relevant limit $g^2 \ll 1$ [8].

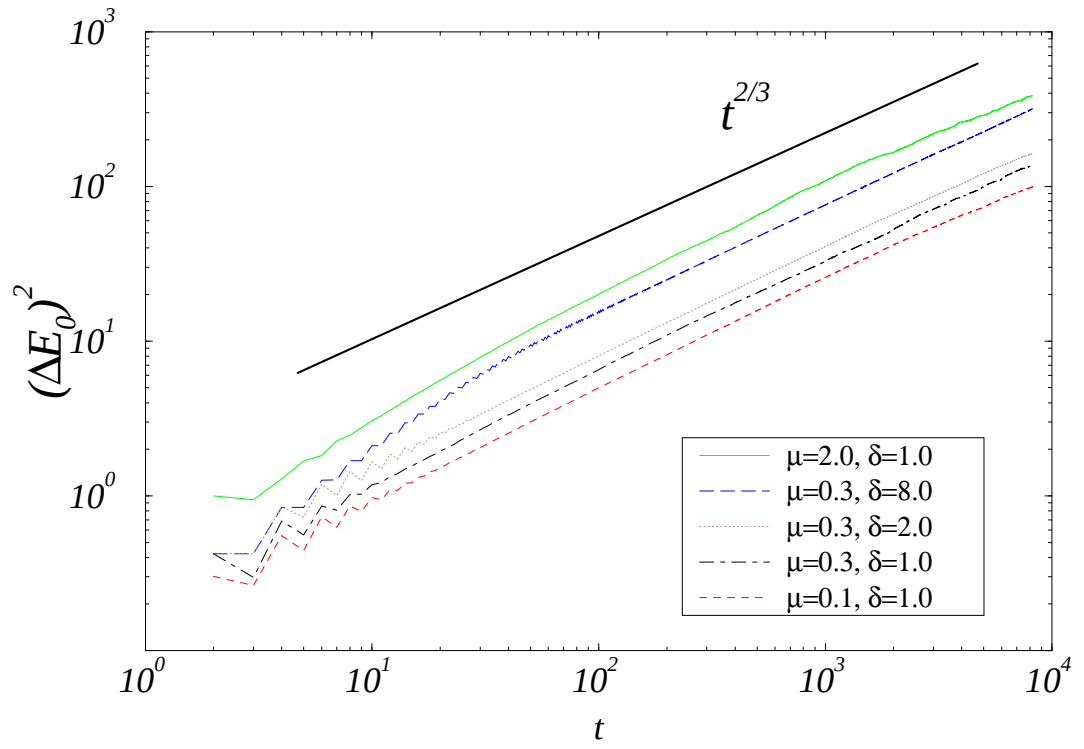

FIG. 4. Energy variations $(\Delta E_0)^2 \equiv \overline{E_0^2(t)} - \overline{E_0(t)}^2 \sim t^{2/3}$ of the optimal alignments. The results are obtained from a sample of 100 pairs of uncorrelated random sequences for each set of the alignment parameters listed.

To study the *high-similarity phase*, we construct sequences $\{P_i\}$ and $\{Q_j\}$ with mutual correlations as they would arise if these sequences developed from a common ancestor sequence (again taken to be an unbiased random sequence) through independent mutations over a time $\theta$. Consider first evolution through point substitutions only: on each sequence, elements are randomly chosen with rate $\Gamma$ and replaced by an unbiased random number $\pm 1$. Then $\{P_i\}$ and $\{Q_j\}$ remain unbiased random sequences and have *mutual correlations* $\overline{P_i Q_j} = p^2 \delta_{i,j}$, where $p = \exp(-\Gamma\theta)$ is the fraction of elements on each sequence that are still unaffected by the mutations after a time $\theta$. It follows that the pairing potential still has the variance (1) (up to an irrelevant [8] term $\sim \delta_{t,t'}\delta_{r,-r'}$), but the average is now $\overline{v_{r,t}} = -J - U\delta_{r,0}$: mutual correlations generate a target path $R(t) = 0$ (along the diagonal of the alignment lattice) that *attracts* the alignment paths with strength $U = gp^2 J$. This interaction reflects the excess number of matches ($P_i = Q_i$) of unmutated daughter elements of the ancestor sequence. It turns out to change the optimal alignment drastically [13]: The path $r_0(t)$ becomes *localized*, i.e., its mean square displacement from the target path $\overline{r_0^2(t)}$ grows no longer superdiffusively, but reaches a finite value $\xi_\perp^2 \equiv \lim_{t\to\infty} \overline{r_0^2(t)}$ for long sequences. In recent years, this localization has been studied carefully in the context of flux pinning in type-II superconductors [13]. It is governed by the competition of the potential energy $\sim U/\xi_\perp$ per unit of $t$ gained from the overlap with the target path, and the random energy cost $V/\xi_\perp$ (with a constant $V \simeq g^2 J$ for $g^2 \ll 1$ [8]) due to the confinement of the alignment path [12]. (This is very similar to the competition of potential and kinetic energy determining the localization of a quantum particle in a potential well.) In the limit $p = 1$ of identical sequences, the optimal alignment is obviously $r_0(t) = 0$; the path is tightly bound to the target. As $p$ decreases, the fluctuations around the target increase, and the overlap decreases. It is found [13] that the path $r_0(t)$ remains localized to the target even for an arbitrarily weak attraction $U > 0$, although the localization length becomes very large for $U \ll V$,

$$\xi_\perp \simeq \exp(bV/U)\,, \tag{3}$$

(with a universal constant $b \approx 2.8$). Thus the fidelity of the optimal alignment, $\xi_\perp^{-1}$, tends to zero with vanishing $U$; this signals a continuous transition to the superdiffusive behavior at $U = 0$.

We now turn to a more realistic model of evolution that includes local insertions and deletions: we generate the sequences $\{P_i\}$ and $\{Q_j\}$ from a common ancestor through point substitutions as before, then we randomly choose a fraction $\tilde{p} \ll 1$ of the sites on each sequence, and insert or delete a random element $\pm 1$ at the chosen sites. This modifies the mutual correlations,

$$\overline{P_i Q_j} = p^2 \delta_{i-j,R(i+j)}\,, \tag{4}$$

and hence the pairing potential $\overline{v_{r,t}} = -J - U\delta_{r,R(t)}$. The target path $R(t)$ is no longer along the diagonal of the alignment lattice, but along the trajectory of a Gaussian random walk with mean square displacement $\overline{(R(t) - R(t'))^2} = \tilde{p}|t - t'|$ (see Fig. 3). Since a fraction $2\tilde{p}$ of the target path involve gaps, the attractive strength of the target is shifted to

$$U = [(gp^2(1-\tilde{p}) - 2\tilde{p})]J\,, \tag{5}$$

and now changes sign at the threshold $p^2 = p_c^2 = 2\tilde{p}/g + O(\tilde{p}^2)$. Above the threshold ($p^2 > p_c^2$), the interaction remains attractive ($U > 0$). In this phase, the optimal alignment path turns out to be localized to the target in much the same way as before [13], with a finite mean square displacement $\xi_\perp^2 \equiv \lim_{t\to\infty} \overline{(r_0(t) - R(t))^2}$ given by (3) and (5) and a finite fidelity $\sim \xi_\perp^{-1}$. As $p^2$ approaches $p_c^2$ from above, the fidelity again tends to zero continuously. For $p^2 < p_c^2$, the interaction becomes repulsive, and the path $r_0(t)$ is again super-diffusive. This is the low-similarity phase, where the optimal alignment does not reflect the mutual correlations: its fidelity is zero. The singular behavior of the fidelity given by (3) can be observed for sequences whose length is larger than the *correlation length* $\xi_\parallel \sim \xi_\perp^{3/2}$; the fidelity for shorter sequences is described in [8].

The concept of an order parameter is useful for the selection of optimal alignment parameters $\mu$ and $\delta$. As follows from Eqs. (3), (5), and $V \simeq g^2 J$, the correlations (4) with given parameters $p$ and $\tilde{p}$ can be detected only for $g > g_c(p,\tilde{p}) = 2\tilde{p}/p^2 + O(\tilde{p}^2)$. They are recovered best if $g$ takes the value $g^\star(p,\tilde{p}) = 4\tilde{p}/p^2 + O(\tilde{p}^2)$ obtained by maximizing $\xi_\perp^{-1}$ for fixed $p$ and $\tilde{p} \ll p^2$. This is a linear condition on $\mu$ and $\delta$.

There are a number of related alignment issues relevant to applications in biology, where our results apply in a similar way; a detailed account will be published elsewhere [8]. (i) For alphabets with a larger character $k$ ($k = 4$ for DNA and $k = 20$ for proteins), the effective strength of attraction of the target path increases, substantially decreasing the localization and correlation lengths. (ii) A higher biological relevance of the alignment is often achieved if gap initiations are penalized by a higher energy than gap extensions. Such refinements lead to alignment paths that have, in addition to their line tension $\gamma$, a finite "bending rigidity". (iii) The alignment of $n$ sequences is described by a directed alignment path $(r_1, \ldots, r_{n-1})(t)$ in $n-1$ transversal dimensions. For $n > 2$, the detection threshold is *increased* to finite $U > 0$, making similarity-detection more difficult. The divergence of the localization length close to the transition is given there by a *power law* instead of Eq. (3). (iv) The same is true even for the alignment of two sequences if there are long-ranged intra-sequence correlations that fall off sufficiently slowly (as may be the case for the non-coding regions of the genome). (v) Unlike the *global* algorithms discussed so far, *local* alignment algorithms match only a contiguous piece of sequence $\{P_i\}$ with a different piece of sequence $\{Q_j\}$; they are appropriate to finding mutual similarities that exist only within these two pieces. It has been noted [5] that the regimes of local and global alignment are separated by a transition line in the space of parameters $(\mu, \delta)$. That transition is quite different from the transition described in this Letter; it can be understood [8] as a boundary-induced critical phenomenon analogous to wetting transitions.

In summary, we have described a unique approach to similarity detection, identifying sequence alignment algorithms with physics problems defined on a lattice with quenched disorder. We show that the successful detection of correlations between sequences depends on the kind of mutations they undergo, as well as on the specific choice of the alignment parameters. This is demonstrated for simple stochastic mutation processes modeling biological evolution. In such systems, correlations can only be detected if their amount exceeds a threshold value; the onset of detection is shown to be a critical phase transition with universal characteristics. Most importantly, it is the *fluctuation statistics* at this transition that determines the fidelity of the optimal alignment. Using that "order parameter", we derive criteria for the optimal choice of alignment parameters given a limited knowledge of the mutation process (in contrast to the current approach of finding these parameters empirically). Conversely, the empirical knowledge of optimal alignment parameters for a given class of proteins can be used to infer the nature of mutations suffered by those proteins.

We thank Michael Zhang for a review of current issues in sequence alignment. TH is supported by an A. P. Sloan Fellowship and an ONR Young Investigator Award. ML is grateful for the hospitality of SUNY at Stony Brook, where much of this work was done.

---